\documentclass[aps,floats,prl,twocolumn,nofootinbib]{revtex4-2}

\usepackage{amssymb}
\usepackage{amsmath}

\usepackage{graphicx}
\usepackage{graphics}
\usepackage[caption=false]{subfig}
\usepackage{dcolumn}
\usepackage{color}
\usepackage{fancyhdr} 
\usepackage{hyperref}

\usepackage[utf8]{inputenc}

\DeclareUnicodeCharacter{00A0}{ }

\usepackage{booktabs}
\usepackage{array}

\makeatletter
\renewcommand*{\fnum@figure}{{\normalfont\bfseries \figurename~\thefigure}}
\renewcommand*{\@caption@fignum@sep}{\textbf{ : }}
\makeatother

\makeatletter
\renewcommand*{\fnum@table}{{\normalfont\bfseries \tablename~\thetable}}
\makeatother

    \newcommand{\be}{\begin{equation}}
  \newcommand{\ee}{\end{equation}}
    \newcommand{\ba}{\begin{align}}
  \newcommand{\ea}{\end{align}}

\newcommand{\vcb}{ v_{\rm cb}  }

\begin{document}

\title{A Standard Ruler at Cosmic Dawn}

\author{Julian B.~Mu\~noz\footnote{Electronic address: \tt julianmunoz@fas.harvard.edu}
} 
\affiliation{Department of Physics, Harvard University, 17 Oxford St., Cambridge, MA 02138}

\date{\today}

\begin{abstract}
The matter in our Universe comes in two flavors: dark and baryonic. 
Of these, only the latter couples to photons, giving rise to the well-known baryon acoustic oscillations and, in the process, generating supersonic relative velocities between dark matter and baryons.
These velocities---imprinted with the acoustic scale in their genesis---impede the formation of the first stars during cosmic dawn ($z\sim 20$), modulating the expected 21-cm signal from this era.
In a companion paper we showed, combining numerical simulations and analytic models,  that this modulation takes the form of robust velocity-induced acoustic oscillations (VAOs), with a well-understood shape that is frozen at recombination, and unaffected by the unknown astrophysics of star formation.
Here we propose using these VAOs as a standard ruler at cosmic dawn.
We find that three years of 21-cm power-spectrum data from the upcoming HERA interferometer should be able to measure the Hubble expansion rate $H(z)$ at $z=15-20$ to percent-level precision, ranging from 0.3\% to 11\% depending on the strength of astrophysical feedback processes and foregrounds.
This would provide a new handle on the expansion rate of our Universe during an otherwise unprobed epoch, opening a window to the mysterious cosmic-dawn era.
\end{abstract}

\maketitle

The 21-cm line of neutral hydrogen is set to revolutionize our understanding of the Universe,  providing access to 
a large cosmic volume unobservable by other probes.
Of particular interest is the cosmic-dawn era, spanning the redshift range $z = 15-30$, which saw the formation of the first stars.
These stars filled the Universe with 
ultraviolet (UV) photons, exciting the hyperfine transition in neutral hydrogen and allowing it to efficiently absorb 21-cm photons from the cosmic microwave background (CMB)~\cite{Wout,Field,Hirata:2005mz}.
In addition, hydrogen was later reheated by the abundant X-rays produced by stellar formation, eventually sourcing 21-cm emission against the CMB~\cite{Furlanetto:2006jb,Pritchard:2011xb,Pritchard:2010pa}.
These two effects allow us to indirectly map the distribution of the first star-forming galaxies during cosmic dawn through the 21-cm hydrogen line.

The first galaxies formed out of matter overdensities at small scales~\cite{loeb2013first,Barkana:2000fd,Springel:2002ux}, where baryons and dark matter (DM) do not behave identically.
After matter-radiation equality the DM started clustering efficiently under its own gravity.
Baryons, on the other hand, were impeded to do so by their interactions with photons, producing the well-known baryon acoustic oscillations (BAOs)~\cite{Peebles:1970ag,Sunyaev:1970eu}.
This discrepancy also generated relative velocities between the two fluids~\cite{Tseliakhovich:2010bj}, which
strongly suppress the formation of the first stars due to their supersonic nature.
Physically, this suppression arises from three sources.
First, large relative velocities damp matter fluctuations at small scales, thus lowering the amount of haloes that can form~\cite{Tseliakhovich:2010bj,Naoz:2011if,Bovy:2012af}.
Second, they allow baryons to stream away from each halo, reducing the amount of gas available for star formation~\cite{Tseliakhovich:2010yw,Dalal:2010yt,Naoz:2012fr}.
Last, they smear overdense gas cores, impeding said gas to cool and form stars~\cite{OLeary:2012gem,Stacy:2010gg,Greif:2011iv,Fialkov:2011iw,Schauer:2018iig,Hirano:2017znw}.

The fluctuations of the DM-baryon relative velocities show marked acoustic oscillations at large scales, due to their BAO origin~\cite{Tseliakhovich:2010bj}.
As a consequence of the three effects outlined above, these oscillations are imprinted into the distribution of the first stars, and thus into the 21-cm power spectrum during cosmic dawn~\cite{Visbal:2012aw,McQuinn:2012rt,Fialkov:2012su,Fialkov:2013jxh,PaperI}.
In Ref.~\cite{PaperI} we showed, via seminumerical simulations with {\tt 21cmvFAST} (a publicly available version of {\tt 21cmFAST}~\cite{Mesinger:2010ne, Greig:2015qca} modified to include streaming velocities), that these unique velocity-induced acoustic oscillations (VAOs) follow a simple analytic shape, which is established at recombination~\cite{Dalal:2010yt,Ali-Haimoud:2013hpa,Munoz:2018jwq}.
This shape is largely impervious to the complex astrophysics of star formation, as the power spectrum of nonlinear functions of the relative velocity is proportional to that of the velocity itself~\cite{PaperI}.
This makes VAOs a powerful probe of acoustic oscillations at high redshifts.

In this {\it Letter} we propose employing a prospective detection of VAOs in the 21-cm power spectrum as a standard ruler to the enigmatic cosmic-dawn era.
The procedure to use VAOs as a standard ruler is similar to regular BAO analyses of galaxy surveys, with two major differences.
First, while matter overdensities are affected by both gravity and the BAOs, relative velocities are only sourced by the latter.
Thus, even though the matter density fluctuates only at the percent level due to the acoustic oscillations~\cite{Ma:1995ey,Eisenstein:1997ik}, 
the relative velocities fluctuate by order unity at acoustic scales~\cite{Tseliakhovich:2010bj}, which
simplifies the task of modeling the VAOs.
Second, galaxy surveys detect roughly isotropically in Fourier space whereas, due to their foreground structure, 21-cm observations are heavily biased towards $k$-modes along the line of sight~\cite{Parsons:2011ew,Morales:2012kf,Datta:2010pk,Parsons:2012qh}.
This hampers a detection of the angular-diameter distance with VAOs, although we will show that it allows for a percent-level measurement of the Hubble expansion rate $H(z)$ of our Universe during cosmic dawn ($z = 15-20$).

\begin{figure}[b]
	\includegraphics[width=0.48\textwidth]{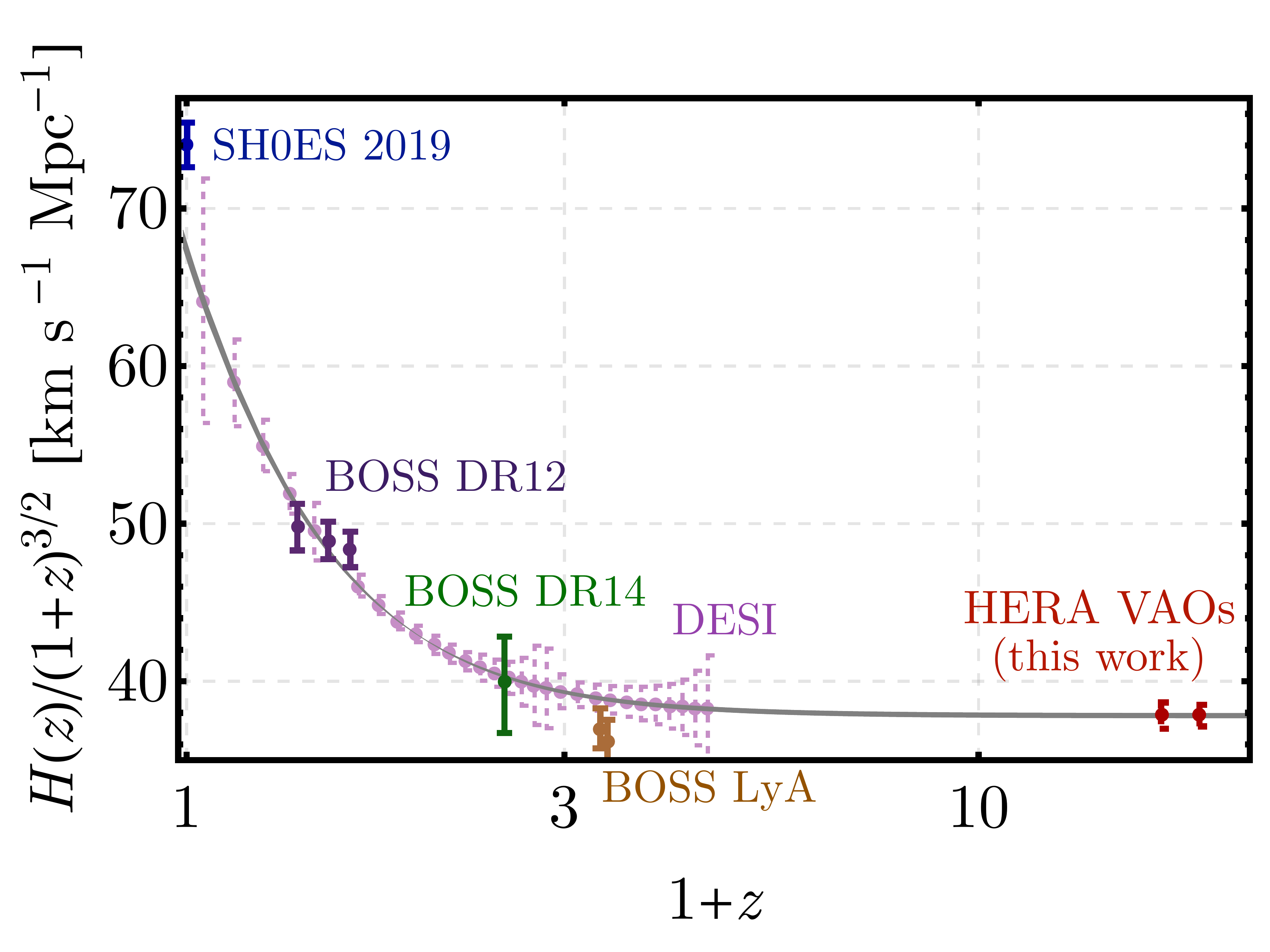}
	\caption{Measurements of the Hubble expansion rate as a function of redshift $z$.
		In dark-purple, green, and brown we show the current constraints from BAO analyses of galaxies, quasars, and the Lyman-$\alpha$ forest, from the Baryon Oscillation Spectroscopic Survey (BOSS)~\cite{Alam:2016hwk,Bourboux:2017cbm,Bautista:2017zgn,Zarrouk:2018vwy}.
		The red points show our projected measurements, obtained through 21-cm observations of the velocity-induced acoustic oscillations (VAOs) with HERA, under the assumptions of moderate foregrounds and regular feedback.
		The gray band represents the uncertainty from current CMB observations, assuming standard cosmology, which
		is in clear tension with the distance-ladder measurement from the Supernova $H_0$ Equation of State (SH0ES) collaboration~\cite{Riess:2019cxk}, shown in blue.
		Finally, the dotted-violet points correspond to forecasted BAO constraints from DESI~\cite{Aghamousa:2016zmz}, which cannot reach the redshifts probed by VAOs.
		In all cases the sound horizon is inferred from Planck CMB data~\cite{Aghanim:2018eyx}.
	}	
	\label{fig:Hzplot}
\end{figure}

Such a measurement would allow us to probe the state of our Universe at an earlier cosmic epoch than any BAO survey.
We illustrate this in Fig.~\ref{fig:Hzplot},
where we show measurements of $H(z)$ through regular BAOs with current datasets~\cite{Alam:2016hwk,Bourboux:2017cbm,Bautista:2017zgn,Zarrouk:2018vwy}, as well as with the future Dark Energy Spectroscopic Instrument (DESI)~\cite{Aghamousa:2016zmz}, 
which in no case can reach redshifts farther than $z \sim 5$.
VAOs, on the other hand, allow us to probe the $z\sim 20$ era, deep into the matter-dominated regime, providing a useful test of exotic physics, such as early dark energy~\cite{Karwal:2016vyq,Hill:2018lfx,Poulin:2018cxd,Agrawal:2019lmo} and decaying DM~\cite{Poulin:2016nat}.
Even within the standard cosmological model, a percent-level measurement of $H(z)$ could help ascertain the origin of the $H_0$ tension between CMB~\cite{Aghanim:2018eyx} and supernovae observables~\cite{Riess:2019cxk,Riess:2016jrr,Rigault:2014kaa} (see also~\cite{Wong:2019kwg}), 
or act as an independent measurement of the acoustic scale $r_d$.
Furthermore, VAOs can  help constrain the masses of the elusive neutrinos, as these contribute to $H(z)$ but not to the growth of fluctuations at small scales~\cite{Lesgourgues:2006nd}.
These examples show the potential of VAOs for the study of cosmology.

VAOs arise from the suppression of the first stellar formation due to the DM-baryon relative velocities ($\vcb$).
These first stars are expected to form in molecular-cooling haloes (with masses $M \lesssim 10^7\,M_\odot$)~\cite{Abel:2001pr,Bromm:2003vv,Haiman:2006si}, where the suppressive effect of $\vcb$ is most sizable.
Nevertheless, the UV background accumulated through gradual stellar emission dissociates molecular hydrogen, so that eventually only atomic-cooling haloes (with $M\gtrsim 10^7\, M_\odot$) form stars~\cite{Oh:2001ex}, lowering the expected VAO amplitude.
This well-known process of Lyman-Werner (LW) feedback on star formation has been extensively studied in the literature~\cite{Machacek:2000us,OShea:2006eik,Wise:2007nb}, albeit always in the absence of streaming velocities.
Therefore, it is not known whether LW feedback acts coherently with the $\vcb$-induced suppression.
To parametrize the large uncertainties in this process, we consider three possible LW-feedback strengths.
Our default scenario is that of ``regular" feedback, where the velocities are assumed to add coherently to the LW feedback, as in Ref.~\cite{Fialkov:2011iw}.
We also consider a case of ``low" feedback strength, with a lower overall impact of the LW flux;
and a ``high" feedback strength, where velocities and LW feedback are incoherent, as in Ref.~\cite{McQuinn:2012rt}.

\begin{figure}[hbtp]
	\includegraphics[width=0.48\textwidth]{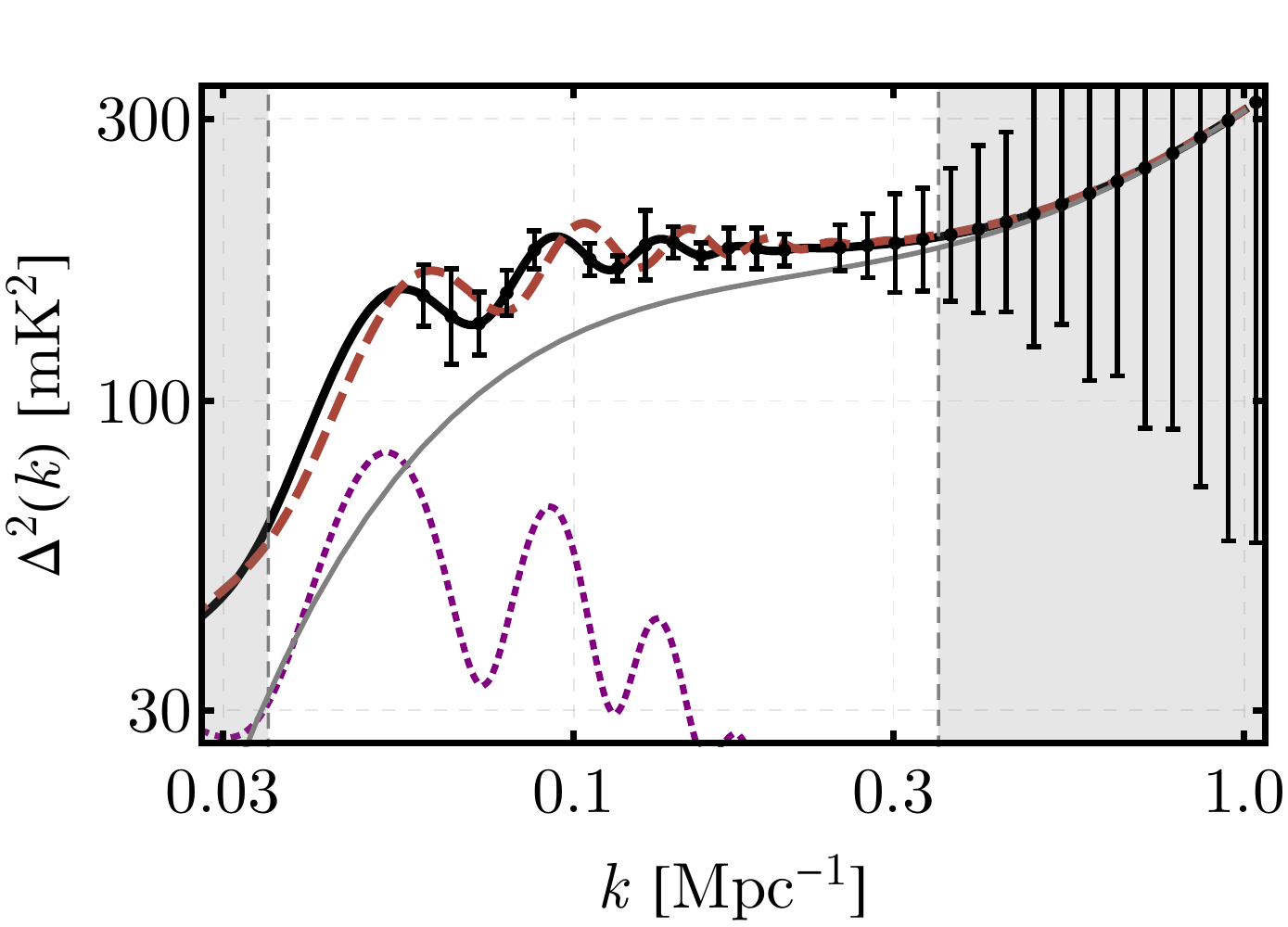}
	\caption{The isotropic 21-cm power spectrum at $z=16.1$ as a function of wavenumber $k$.
	The black-thick line shows the result of our simulations with {\tt 21cmvFAST}---assuming regular feedback strength---with clearly marked velocity-induced acoustic oscillations (VAOs).  
	We can decompose the total signal into a VAO-only component (the dotted-purple line) and a smooth function that we marginalize over (the thin gray line), as described in Eq.~\eqref{eq:model}.
	The VAO peaks get shifted through the Alcock-Paczynsky effect, which we illustrate with the red-dashed line, where we assumed a Hubble parameter 10\% lower than our fiducial at $z=16.1$. 
	The shown error bars have been obtained with {\tt 21cmSense}, and correspond to 540 total days of observation with HERA under the assumption of moderate foregrounds.
	}	
	\label{fig:P21k}
\end{figure}

We will use the 21-cm hydrogen line to indirectly probe the distribution of star-forming galaxies at cosmic dawn.
Our observable will be the dimensionless 21-cm power spectrum $\Delta^2(k)$,
which we obtain via semi-numerical simulations with {\tt 21cmvFAST}\footnote{\url{ https://github.com/JulianBMunoz/21cmvFAST}. See also \\ \url{https://github.com/andreimesinger/21cmFAST}
 }~\cite{PaperI}.
We show the predicted 21-cm power spectrum at $z=16.1$ in Fig.~\ref{fig:P21k} for our default scenario, where the power is generated by the inhomogeneous X-ray heating of the hydrogen gas.
Here the relative velocities produce fluctuations in the X-ray flux from the first galaxies, generating clearly visible VAOs in the 21-cm power spectrum.

The acoustic origin of the VAOs imprints the scale $r_d$ onto  the observable 21-cm signal, where $r_d \approx 150$ Mpc is the comoving sound horizon at the baryon drag era ($z_d \approx 1060$)~\cite{Aghanim:2018eyx}.
This is apparent, for instance, from the separation between the VAO maxima of $\Delta k = 2\pi/r_{d} \approx 0.04$ Mpc$^{-1}$ in Fig.~\ref{fig:P21k}.
Therefore, the presence of VAOs provides us with a well-known distance scale at cosmic dawn.
We use it as a standard ruler by performing an Alcock-Paczynsky (AP) test on our data~\cite{Alcock:1979mp}:
A feature at some wavenumber $(\mathbf k_\perp, k_{||})$, where the subscripts $\perp$ and $||$ represent the perpendicular and line-of-sight (LoS) directions, is shifted to $(\mathbf k_\perp/\alpha_\perp, k_{||}/\alpha_{||})$ when assuming the wrong fiducial cosmology. 
The two AP parameters are
\be
\alpha_{||} = \dfrac{H^{\rm fid}(z) r_{d}^{\rm fid}}{H(z) r_{d}}   \quad {\rm and} \quad 
\alpha_\perp = \dfrac{D_A(z) r_{d}^{\rm fid} }{D_A^{\rm fid}(z) r_{d}},
\ee
where $H$ is the Hubble expansion rate and  $D_A$ the angular-diameter distance, and the superscript ``fid" stands for fiducial.
Then, by searching for shifts in the VAO peaks we can constrain both $\alpha_{||}$ and $\alpha_\perp$, thus measuring $H(z)$ and $D_A(z)$ at the observation redshift.
As an example, in Fig.~\ref{fig:P21k} we show the 21-cm power spectrum that would be inferred if $H(z=16.1)$ were 10\% smaller than our fiducial value (corresponding to $\alpha_{||}=1.1$), which would clearly shift the VAO peaks to smaller scales.
Note that here we work in $k$-space for simplicity, although it would be equivalent to work directly in visibility space~\cite{Mao:2008ug}.

We make use of two facts that vastly simplify the use of VAOs as a standard ruler.
First, to a good approximation the fluctuations of any smooth function of $\vcb$ are proportional to
\be
\delta_{v^2} \equiv \sqrt{\dfrac{3}{2}} \left[(\vcb/v_{\rm rms})^2-1\right],
\ee 
where $v_{\rm rms}$ is the root-mean-square value of $\vcb$.
We denote the power spectrum of this quantity as $\Delta^2_{v^2}(k)$, which defines the shape of the VAOs~\cite{PaperI}.
Second, VAOs are uncorrelated with the usual 21-cm fluctuations sourced by overdensities~\cite{Dalal:2010yt,Ali-Haimoud:2013hpa,Munoz:2018jwq}, and so they can be linearly added to the usual (no-$\vcb$) 21-cm power spectrum to obtain the total signal.
Consequently, we model the 21-cm power spectrum as
\be
\Delta^2_{\rm model}(k,z) = A_{\rm vel}(z) \Delta^2_{v^2}(k) W^2(k,z) + \mathcal P_n(k,z),
\label{eq:model}
\ee
where $A_{\rm vel}$ (with units of mK$^2$) is the VAO amplitude, and $W(k,z)$ is a window function accounting for the nonlocal propagation of X-ray photons from the first stars~\cite{Dalal:2010yt} (which only produces a modest suppression in power within the $k$ range of interest~\cite{PaperI}).
We use an $n$-th order polynomial, 
\be
\mathcal P_n(k,z) = \exp \left ( \sum_{j=0}^n c_j(z) \left[\log(k)\right]^j \right),
\ee
to parametrize the smooth  nonoscillatory part of the 21-cm signal, where $c_j$ are nuisance parameters to be determined from data.
We show in Fig.~\ref{fig:P21k} the VAO-only contribution to the 21-cm power spectrum, with the expected large acoustic oscillations at $k\sim 0.1$ Mpc$^{-1}$, as well as the smooth $\mathcal P_n$ component, both at $z=16.1$.
We note, in passing, that the power at large scales ($k\lesssim 0.3$ Mpc$^{-1}$) increases when including VAOs.

In all of our simulations the X-ray heating era roughly starts at $z\approx20$ (where the global 21-cm signal is minimum), and lasts until $z\approx15$ (where the global signal crosses zero, transitioning into emission), so in this work we consider two redshift bins, centered at $z=16$ and $z=18$, encompassing a $\Delta z=1$ above and below their centers.
These bins are only meant to be illustrative, since the heating era can be shifted to earlier or later times by altering the (unconstrained) X-ray luminosity, as further explored in Ref.~\cite{PaperI}, which however does not alter the main results of this work.

Whether we can detect the predicted shift in the VAO peaks---and thus measure $H(z)$---depends not only on the sensitivity of the experiment at hand, but also on the severity of the foregrounds.
These are expected to contaminate a large region of the observable Fourier space, usually termed the ``wedge"~\cite{Parsons:2011ew,Morales:2012kf,Datta:2010pk,Parsons:2012qh}, which is deemed irretrievable for cosmological studies (as the foregrounds outweigh the cosmic signal roughly $10^8$ to 1~\cite{Pober:2013ig,Liu:2014yxa}).
We follow Refs.~\cite{Pober2013,Pober:2013jna} in parametrizing the extent of the foreground wedge by assuming that all wavenumbers with $k_{||}$ below
\be
k_{||}^{\rm min} = a + b(z) k_\perp
\label{eq:wedge}
\ee
are contaminated, where $b(z)\approx 6$ accounts for the chromaticity of the antennae, and $a$ is a constant superhorizon buffer~\cite{Parsons:2012qh}.
Given the large foreground uncertainties, we study three cases based on (but not identical to) those of Ref.~\cite{Pober:2013jna}. 
In the pessimistic- and moderate-foreground cases we take the usual $b(z)$ determined by the horizon limit, with buffers of $a=\{0.1,0.05\}\,h\,\rm Mpc^{-1}$, respectively.
In the optimistic case we set $a=0$ and $b (\approx 1)$ given by the primary beam~\cite{Pober:2013jna}, and in all cases we assume that different baselines can be added coherently.

For concreteness we focus on the Hydrogen Epoch of Reionization Array (HERA)\footnote{\url {reionization.org} }~\cite{DeBoer:2016tnn}, for which we obtain sensitivity curves using the publicly available code {\tt 21cmSense}\footnote{\url{https://github.com/jpober/21cmSense}}~\cite{Pober2013,Pober:2013jna}, with two minor modifications.
First, we bin the $k$-modes logarithmically, instead of linearly, to better resolve the VAOs.
Second, at each redshift we split the available data between different bandwidths.
This is designed to observe a larger amount of wavenumbers, as the foreground wedge only allows a small range of $k_\perp$ modes to be observable around each $k_{||}$, and the fast Fourier transform (FFT) performed within each bandwidth $B$ determines which $k_{||}$ modes can be observed (through $k_{||} \propto N/B$ for integers $N$).
While this splitting will allow us to better probe the shape of the 21-cm power spectrum (and thus more clearly characterize the oscillations), it will also reduce the sensitivity at each individual wavenumber.
We only consider bandwidths below 8 MHz, to keep each redshift slice roughly in the co-evaluation regime. 
Thus, we use three bands at each redshift, with widths $B=\{6,7,8\}$ MHz both for $\nu =83$ MHz (corresponding to $z=16$) and $\nu =75$ MHz (corresponding to $z=18$), where each band uses data from 180 observation days (totaling 540 days).
The resulting HERA error bars are shown in Fig.~\ref{fig:P21k}.

Given our model, from Eq.~\eqref{eq:model}; the mock data $\Delta^2_{\rm data}$, from {\tt 21cmvFAST}; and its error bars $\delta \Delta^2(k)$, from projected HERA observations, we define our likelihood $\mathcal L$ at each redshift bin through
\be
-\log \mathcal L = \dfrac{1}{2} \sum_{k \rm-bins} \dfrac{\left[\Delta^2_{\rm data}(k) - \Delta^2_{\rm model}(k;\mathbf p) \right]^2}{ [\delta\Delta^2(k)]^2},
\ee
where $\mathbf p$ is a parameter vector that we will specify.
We employ data in the range $k=(0.05-0.5)\,h\,\rm Mpc^{-1}$ (as lower wavenumbers are contaminated, and higher ones do not show VAOs), which we divide into 
$k-$bins, and sample the likelihood with the Python package {\tt emcee}~\cite{ForemanMackey:2012ig}.
Note that, as a consequence of the foreground wedge, we will likely only be able to measure wavenumbers with $k_{||} \gg k_\perp$.
In that case, we can disregard variations in the AP parameter $\alpha_\perp$, as they are negligible compared to those in the LoS parameter $\alpha_{||}\propto [H(z) r_d]^{-1}$.
Under this approximation, our parameter vector for each redshift bin will be $\mathbf p = \{\alpha_{||}, A_{\rm vel}, \mathbf c \}$, 
where $\mathbf c$ is a vector of length $n+1$, containing the (nuisance) coefficients of the smooth polynomial $\mathcal P_n$.
We impose a prior of $0.8 \leq \alpha_{||} \leq 1.2$ to avoid unphysical values~\cite{Gil-Marin:2015nqa} (such as $\alpha_{||} = 0$), as well as $0\leq A_{\rm vel}^{(i)}\leq 10^3\,\rm mK^2$ and $-20 \leq c_{j}\leq 20$, which are broad enough to fit the 21-cm power spectrum in all of our simulations.
Additionally, we determine the order $n$ of the nonoscillatory polynomial ($\mathcal P_n$) by finding the first coefficient $c_{n+1}$ that is consistent with zero, given our predicted uncertainties. 
These depend on the foreground severity, and we find that for the case of pessimistic and moderate foregrounds $n=1$ suffices to properly fit the non-VAO power spectrum within the $k$ range that we are interested in, whereas when considering optimistic foregrounds the expected noise level is lower, and $n=2$ is required.

\begin{table}[hbtp!]
	\noindent\begin{tabular}{ c | c  c  c }	
		\multicolumn{1}{c}{} 	&\multicolumn{3}{c}{Foregrounds}\\
		\cline{2-4}
		Feedback strength  &  Pessimistic & Moderate &  Optimistic \\
		\hline
		High & $-$ & 8.5\% & 6.9\%  \\
		
		Regular  & 11\%  & 2.2\% & 0.6\% \\
		
		Low  & 3.1\%  &  1.1\% & 0.4\% \\
\hline
\hline		
	\end{tabular}
	\noindent\begin{tabular}{ c | c  c  c }	
	Feedback strength  &  Pessimistic & Moderate &  Optimistic \\
	\hline
	High & $-$ & 8.9\% & 2.8\%  \\
	
	Regular  & 4.1\%  & 1.8\% & 0.7\% \\
	
	Low  & 1.7\%  &  0.7\% & 0.3\% \\
	
\end{tabular}
	\caption{Projected relative errors on $H(z)r_d$ (at 68\% C.L.) at $z=16$ (\emph{top}) and $18$ (\emph{bottom}), under our different foreground assumptions and feedback models, as detailed in the main text, in all cases with 540 days of HERA observation.
	The high-feedback pessimistic-foreground case does not have enough sensitivity to reach a detection at any precision.
	}
	\label{tab:errorbars}
\end{table}

We show our forecasted marginalized sensitivities to $H(z) r_d$ (obtained from $\alpha_{||}$) in Table~\ref{tab:errorbars}, both at $z=16$ and $18$, for each of our feedback and foreground assumptions.
This Table is the main result of this work.
In all cases but one it is possible to detect the VAOs with enough significance to obtain a measurement of $H(z) r_d$, with precision ranging from sub-percent (competing with current determinations of $r_{d}$ from Planck~\cite{Aghanim:2018eyx}) under optimistic assumptions, to 11\% for the  most pessimistic cases.
Focusing on our default scenario of moderate foregrounds and regular feedback, we put our VAO projections in context by comparing them with regular BAO measurements of $H(z)$ in Fig.~\ref{fig:Hzplot}, where the unique large-$z$ reach of the VAOs is apparent.
In all of our results we have assumed 540 days of HERA data, albeit in most cases a third of that is sufficient to detect the VAOs, which, however degrades the precision on $H(z) r_d$ by $\sim 50\%$ with respect to Table~\ref{tab:errorbars}.

\begin{figure}[t]
	\includegraphics[width=0.5\textwidth]{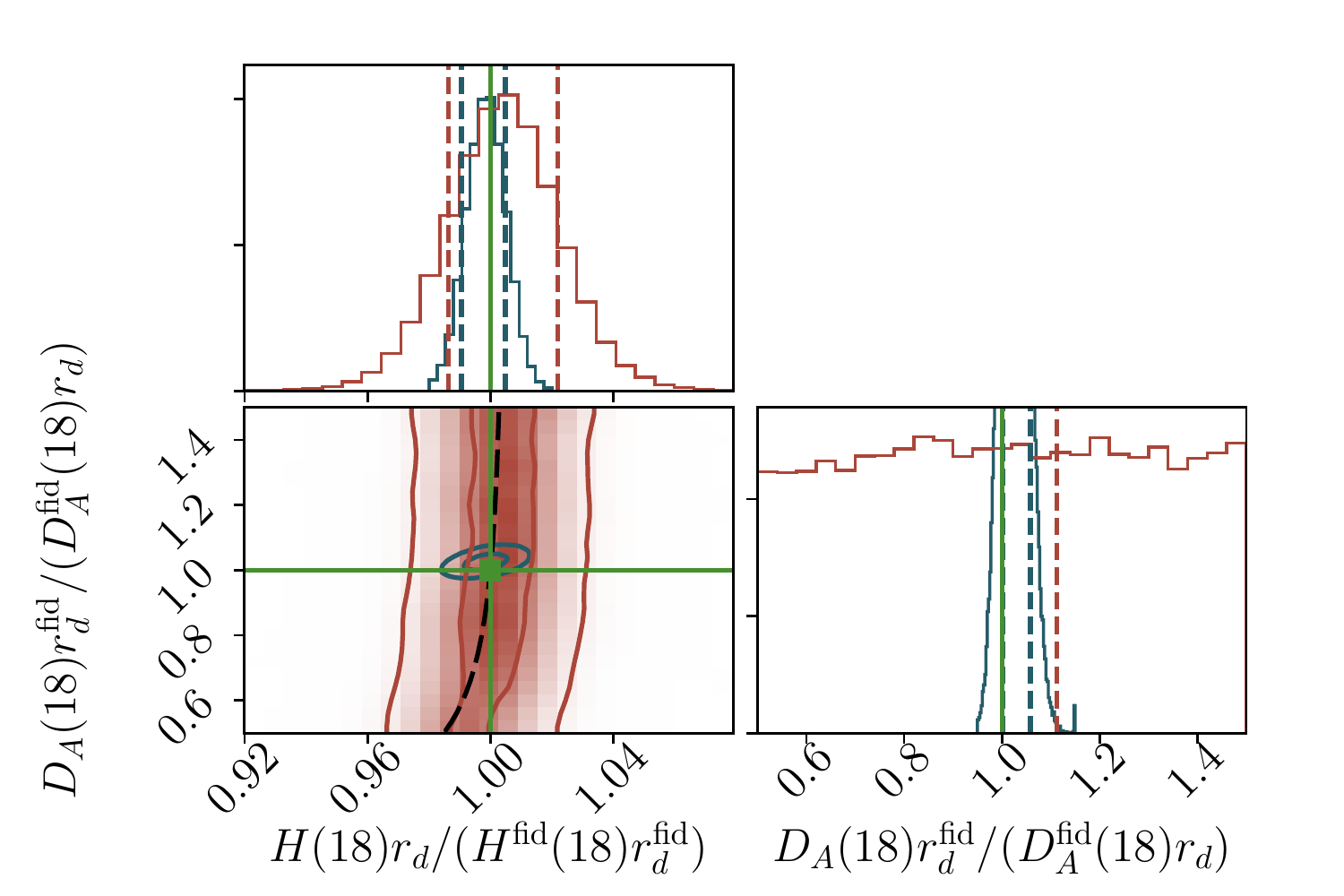}
	\caption{Projected 1- and 2-$\sigma$ confidence contours  for $H(z) r_d$ and $D_A(z)/r_d$ at $z=18$, normalized to their fiducial values, under two different foreground models: moderate (red) and optimistic (blue), assuming regular feedback in both cases.
		The black-dashed degeneracy line has a fixed isotropic AP parameter~\cite{Ross:2015mga}, and the green lines mark the fiducial values of unity for reference.
	}	
	\label{fig:ellipse2D}
\end{figure}

Before concluding, let us briefly study what constraints could be placed on $D_A$ through a fully anisotropic AP test of the VAOs.
For this, we extend our likelihood to depend on $k_\perp$ and $k_{||}$ independently, where now our vector parameter $\mathbf p$ contains both $\alpha_{||}$ and $\alpha_\perp$,
and given the typical small range of $k_\perp$ observable we assume that both the signal and our model are not explicitly anisotropic (see, however, Refs.~\cite{Datta:2011hv,Greig:2018hja}).
We obtain the noise $\delta[\Delta^2(k_\perp,k_{||})]$ in each bin through a modified version of {\tt 21cmSense}, and
show our projected confidence ellipses for $D_A/r_d$ and $H r_d$ (both normalized to their fiducial values) at $z=18$ in Fig.~\ref{fig:ellipse2D}, assuming regular feedback, using the {\tt corner} package~\cite{corner}.
From this figure we see how, as predicted, moderate foregrounds do not allow for a meaningful measurement of the angular-diameter distance.
Indeed, we find $D_A(18) r_d^{\rm fid}/[D_A^{\rm fid}(18) r_d] = 2.7\pm1.6$ at 68\% C.~L. for this case, although the relative error projected for $H(z) r_d$ is 1.8\%, unaffected by the inclusion of $D_A$.
The situation is more promising under optimistic foregrounds, where we can measure $D_A/r_d$ to 2.8\% and $H r_d$ to 0.7\% precision, with small correlation between them. A measurement of $D_A(z\sim18)$ would place strong constraints on both the curvature of our Universe and the evolution of dark energy~\cite{Kovetz:2017agg},
showing  that foreground removal is critical to fully exploit the information in 21-cm observables during cosmic dawn, akin to lower-$z$ analyses~\cite{Obuljen:2017jiy}.

In addition to VAOs, the 21-cm signal is affected by the ``regular" (density-induced) BAOs~\cite{Barkana:2005xu,Barkana:2005nr,Chang:2007xk,Wyithe:2007rq,Kashlinsky:2015pna,Obuljen:2017jiy,Ansari:2018ury}, which however are much smaller in amplitude.
Although BAOs are a promising standard ruler for 21-cm surveys at lower redshifts~\cite{Kovetz:2017agg,Barkana:2004zy,Pober:2014lva}, 
the complicated mapping between densities and 21-cm signal at cosmic dawn hinders their use during this era.
The VAOs sidestep this issue, as they produce large oscillations with a well-understood shape~\cite{PaperI}.
Moreover, while here we have only studied the epoch of X-ray heating, a similar analysis could be carried out during the preceding Lyman-$\alpha$ coupling era, which lasts from $z=20-28$ for our fiducial parameters.

There are some caveats about our analysis worth mentioning.
We have considered a broad range of feedback and foreground assumptions, and found that VAOs are observable in almost all cases.
Nevertheless, there might be additional sources of feedback (such as mechanical or radiative~\cite{Barkana:1999apa,Mesinger:2008ze,Okamoto:2008sn,Springel:2002ux}) that conspire to hide the VAO signal,
preventing the formation of even atomic-cooling haloes during cosmic dawn.
Similarly, the spectrum of the first X-ray sources can affect the detectability of VAOs, as higher-energy photons have longer mean-free paths~\cite{Pacucci:2014wwa,Fialkov:2014kta}, damping small-scale fluctuations~\cite{PaperI,Dalal:2010yt,Fialkov:2015fua}.
We have tested our method assuming a higher X-ray cutoff energy $E_0=0.5$ keV, as described in Ref.~\cite{PaperI}, and found very similar results to those with our fiducial cutoff at $E_0=0.2$ keV (both with and without updating the window function in Eq.~\eqref{eq:model}), as the damping of VAOs is compensated by a lower nonoscillatory signal.
Nonetheless, for cutoffs above $1$ keV  it might become impossible to detect the VAOs, and thus to measure $H(z)r_d$.
We emphasize, however, that different astrophysical effects
can alter the observability of the VAOs but not their unique acoustic shape.
So, while detecting VAOs is not guaranteed, such a detection would provide a robust standard ruler at cosmic dawn.

In summary, the DM-baryon relative velocities are predicted to leave striking VAOs on the 21-cm power spectrum at $z =15- 20$. 
We have shown how, 
by using the acoustic scale imprinted by these 
VAOs as a standard ruler, 
the HERA interferometer should be able to measure the cosmological expansion rate at cosmic dawn,
casting light onto the properties of our Universe during this mysterious era.

\acknowledgements
 
It is our pleasure to thank Yacine Ali-Ha\"imoud, Marc Kamionkowski, David Pinner, and Matias Zaldarriaga for enlightening discussions; Jonathan Pober for help with the HERA noise; and Cora Dvorkin, Abraham Loeb, and especially Ely Kovetz for comments on a previous version of this manuscript.
Some computations in this Letter were run on the Odyssey cluster supported by the FAS Division of Science, Research Computing Group at Harvard University.
This work was supported by the Department of Energy (DOE) grant DE-SC0019018.

\bibliography{21cmBAO}

\end{document}